\documentclass{elsart}

\usepackage{epsfig}

\makeatletter
\newif\if@restonecol
\makeatother

\usepackage{algorithm2e}

\usepackage{amssymb}

\begin{document}

\begin{frontmatter}



\title{An efficient algorithm for granular dynamics simulations with 
complex-shaped objects}


\author{Fernando Alonso-Marroqu\'{\i}n and Yucang Wang} 

\address{ESSCC, The University of Queensland, Qld. 4072, Brisbane, Australia}

\begin{abstract}

The most difficult aspect of the realistic modeling of granular materials is how
to capture the real shape of the particles. Here we present a method to simulate 
granular materials with complex-shaped particles. The particle shape is represented 
by the classical  concept of a Minkowski sum,  which permits the representation 
of complex shapes  without the need to define  the object as a composite of  
spherical or convex  particles.  A well defined interaction force between these  
bodies is derived.  The algorithm for  identification of neighbor particles 
reduces force calculations to  $O(N)$, where  $N$ is the number of particles. 
The algorithm is   much more  efficient, accurate and easier to implement than other 
models.. We investigate the existence of a statistical  equilibrium in granular systems 
with circular non-spherical particles in the collisional. regime. We also investigate the 
limit state of dissipative granular materials using  biaxial test simulations. The results 
agree with the classical assumption of the statistical mechanics for non-dissipative systems, 
and the critical state theory of soils mechanics for dissipative granular materials. 
\end{abstract}

\end{frontmatter}


\section{Introduction}

Rapid advances in computer simulations have led to many new developments in 
the modeling of particulate systems.  
These systems represent different real physical systems at different scales, such as  
the small  scale of liquid crystals \cite{Pelzl99}, geological scales of snow and debris 
flow, and the astronomical  scales of planetary rings or dynamics evolution of precursors 
of planets \cite{poeschel00}. Although particle shape plays an important role, most  
theoretical  and numerical developments have been restricted to particles with spherical 
or circular shape. These simplification lead in some cases to unrealistic properties.
In collisional non-dissipative systems, spherical (or circular) particles can not 
exchange angular momentum, so that the system cannot explore all the phase space during
the evolution.  In dissipative granular systems such as sand piles or fault gouges,
disks of spheres tend to roll more easily than non spherical particles, leading to
unrealistic angles of repose and bulk friction coefficients.

Three different approach has been presented to model the real shape of particulate
materials. In the first approach the shape is represented as a scalar funcions. This
model allows to represent especific shapes, such as ellipses \cite{ting1993ebd},
ellipsoids \cite{lin1997tdd} and superquadric \cite{mustoe2001mfa}.
The main drawback of those methods is that the calculations required in the contact
force are much more expensive than in spherical (or circular) particles. In the second
approach the non-spherical particle is represented ad  aggregates or clumps of disks and 
spheres bonded together \cite{cheng2003des,mcdowell96}. In this approach  crushing and 
fracture of aggregates  can be easily modeled. The disadvantage is that this method requires 
a larger number of particles.

The third approach is represent the complex shape using polygons in 2D,
\cite{alonso02a,matuttis00,mirghasemi02}  or polyhedrons in 3D \cite{mirtich98,cundall1988ftd}
The most difficult aspect for the simulations of these objects is the handling of contact interactions. 
Nowadays, the interaction is resolved by decomposing them in convex pieces, and applying penalty
 methods, impulse-based methods or dynamic constraints in the interaction between these pieces.  
Impulse-based methods (also called even-driven methods) allow real-time simulations, but they cannot 
handle permanent or lasting contacts \cite{mirtich98}. On the other hand, constraint methods 
(or contact dynamics methods) can handle resting contacts with infinite stiffness, but simulations 
are computationally expensive and lead in some cases to indeterminacy in the solution of contact 
forces . This indeterminacy is removed by using penalty methods, where the bodies are 
allowed to interpenetrate each other and the force is calculated in terms of their overlap. 
However, until very recently the determination of such contact force has been heuristic and 
lacks physical correctness, because the interactions do not comply with energy conservation 
\cite{poeschel04}

We propose a solution to this problem in 2D based in the mathematical concept of 
spheropolygons.  These objects are generated from the Minkowski  sum of a polygon with a disk, 
which is nothing more than the object resulting from sweeping 
a disk around the polygon. The 3D counterpart of these objects are the spheropolyhedrons, which
were recently introduced by Pournin and Liebling for the simulations of granular media.
This simple concept can be used to  generate very complex shapes, including non-convex bodies, 
without the need to decompose them  into spherical or convex parts. Here we introduce 
an efficient method to calculate dissipative and non-dissipative interaction between
spheropolygons. We probe also  that the model complies with the statistical mechanical 
principles and the physical laws of a  conservative system. Since the code can simulate 
particles a wide range of shapes, it can be used  to investigate the effect of the aspect 
ratio, angularity and non-convexity of the particles on  granular flow. The relevance of 
this investigation is demonstrated in the stress-strain response  in biaxial test bellow, 
where we show that  the system reach the critical state of soil mechanics, where the particle 
shape strongly affect the material  properties of the granular  media.

\section{Model}

Systems with different particle shapes are modeled using the concept of the Minkowski
sum of a polygon (or polyline) with a disk. This mathematical operation is explained
in Sub-Section \ref{Minkowski sum}. Sub-section \ref{mass} leads with the  numerical 
calculation of mass properties . The interaction between the particles is obtained from 
the individual interaction  between each vertex of one polygon and each edge of another,
as explained in Sub-Section \ref{interaction}. In Sub-Section \ref{efficient} we present
a method to redocue the number
of floating point operations used to calculate interaction forces is drastically reduced 
by using a neighbor list and a contact list for each pair of neighbor particles. In
Sub-Section \ref{time integration} we present the algorith used in each time step of 
the dynamic simulation.

\subsection{Minkowski sum}
\label{Minkowski sum}
Given  two sets of points P and Q in an Euclidean space, their Minkowski sum is given  by   
$P+Q=\{\vec x+ \vec y~|~ \vec x \in P,~ \vec y \in Q \}$. 
This operation is geometrically equivalent to the sweeping of one set around 
the profile of the other without changing the relative orientation.  A special 
case is the sum of a polygon with a disk, which is defined here as 
spheropolygon. Other examples of a Minkowski sum are the spherocyllinder 
(sphere + line segment) \cite{pournin05a}, the spherosimplex  (sphere+simplex) 
\cite{pournin05b} and the spheropolyhedron (sphere+polyhedron) \cite{pournin05c},
which are used in simulations of  particulate systems.

The main advantage of the spheropolygons is that they allow us to represent any
shape in 2D, from rounded to angular particles, and from convex to non-convex
shapes. As we will see in Sub-Section \ref{interaction} The Minkowsky sum does
not need to be explicitely calculated to determine the particle interaction. 
The calculation of the mass properties, however, are calculated numerically, but
this does not affect much the simulation time because the calculations are done
only a the beginning of the simulations.

\begin{figure}
  \begin{center}
    \epsfig{file=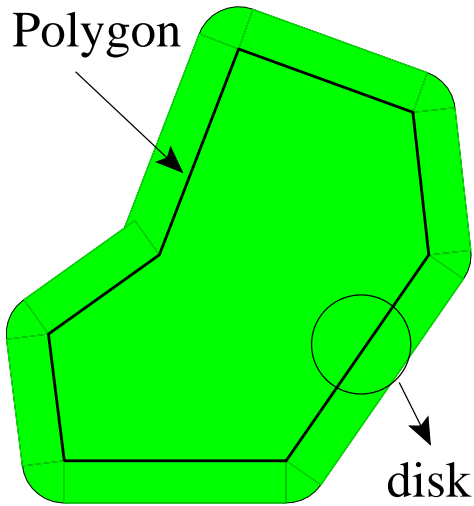, width=0.45\linewidth}
    \caption{Minkowski sum of a polygon with a disk.}
   \label{fig:Minkowski cows}
  \end{center}
\end{figure}

\subsection{Mass properties}
\label{mass}

Before calculating the mass, center of mass and moment of inertica of the spheropolygons 
we need to introduce some  useful  concepts. 
Given a spheropolygon $SP = P+S$, the polygon $P$ will be called the polygon-base; and 
the radius $r$ of the disk $S$ sphero-radius. The distance 
$d(\vec X, P)$ from a test point $\vec X$ to the polygon-base is defined as follows: 
If the point is outside the polygon, it is given by the minimum
distance between the point and the edges of the polygon; If the point is 
inside the polygon, we let $d(\vec X, P) = 0$. Finally, the point $\vec X$ is 
inside of the spheropolygon when it satisfies $d(\vec X, P) < r$.
 
The point-inside-spheropolygon test combined with a basic Monte Carlo method 
is used to evaluate the integral expressions for mass, center of mass and the 
moment of inertia. The numerical integration is performed by taking a 
quasi-random set of points $\vec X_i$ uniformly distributed in a rectangular 
box containing the object. Then the integral over the area enclosed by the 
spheropolygon of any function $f(\vec X)$ is calculated as:

\begin{equation}
\mathbf{M}  = \int_{SP}{f(\vec X) da} \approx \frac{A_{box}}{N_p}\sum^{N_p}_{i=1}{\chi(\vec{X_i})f(\vec{X_i})}.
\label{eq:integral}
\end{equation}

\noindent
$A_{box}$ is the area of the rectangular box, $\vec X_i$ is a quasi-random 
point  inside $A_{box}$, and $N_p=1.6 \times 10^4$ is the number of points.  $\chi(\vec X)$  is the 
characteristic function, which returns one if $\vec X$ is inside the 
spheropolygon  and zero otherwise.  
Replacing $f(\vec X)$ by $\sigma$, $\sigma\vec X$ $\sigma||\vec X||^2$ results in 
$\mathbf{M} = m, m \vec r, I+m||\vec r||^2$ respectively, where $\sigma$ is the density, and  $m$, $\vec r$ and $I$ 
are the mass, center  of mass and moment of inertia.

\subsection{Interaction force}
\label{interaction}

To solve the interaction between spheropolygons we consider all vertex-edge
distances between the polygons base. we consider two spheropolygons $SP_i$ and 
$SP_j$ with their respective polygons base $P_i$ and $P_j$ and sphero-radii 
$r_i$ and $r_j$. Each polygon is defined by the set of vertices $\{V_i\}$ and 
edges $\{E_j\}$.  The overlapping length between each pair of vertex-edge
$(V_i,E_j)$ is defined as

\begin{equation}
\delta(V,E)=\langle r_i+r_j-d(V,E) \rangle,
\label{eq:overlap}
\end{equation}

\noindent
where $d(X,E)= ||\vec Y-\vec X||$
is the Euclidean distance from the vertex $V$ to the segment $E$. Here 
$\vec X$ is the position of the vertex $V$ and $\vec Y$ is its closest point 
on the edge $E$. The ramp function  $\langle x\rangle$ returns $x$ if $x>0$ 
and zero otherwise. The overlapping length in Eq.~(\ref{eq:overlap})
is equivalent to the interpenetration between the disks of radii 
$r_i$ and $r_j$ centered on $\vec X$ and $\vec Y$.

The force $\vec F_{ij}$ acting on particle $i$ 
by the particle $j$ is defined by:

\begin{equation}
\vec F_{ij}=-\vec F_{ji}=  \sum_{V_i E_j}{\vec F(V_i,E_j)}
               +\sum_{V_j E_i}{\vec F(V_j,E_i)},
\label{eq:contact force}
\end{equation}

\noindent
where $F(V,E)$ represent the force between the vertex $V$ and the edge $E$. 
if the vertex-edge pair do not overlap, $F(V,E)=0$. Different of
vertex-edge forces can be included in the modelL linear dashpots,
non-linear Hertzian laws, dissipative viscous forces proportional to the
relative normal and tangential velocities, sliding friction, etc
The force of Eq. \ref{eq:contact force}  is applied to each particle  in the middle 
point of the overlap region between the vertex and the edge:

\begin{equation}
\vec R(V,E) = \vec X + (r_i + \frac{1}{2} \delta(V,E)) 
\frac{\vec X-\vec Y}{||\vec Y-\vec X||},
\end{equation}

\noindent
so that the resulting torque on particle $i$ given by $j$ is

\begin{equation}
\begin{array}{clcr}
\tau_{ij} & = & \sum_{V_i E_j}{(\vec R(V_i,E_j)  - \vec r_i)
                        \times \vec F(E_i,V_j) } \\
          &+ &\sum_{V_j E_i}{ (\vec R(V_j,E_i) - \vec r_i)
                        \times \vec F(E_j,V_i) }, 
\end{array}
\label{eq:torque}
\end{equation}

\noindent
where $\vec r_i $  is the center of mass of particle $i$. 

The evolution of  $\vec r_i$ and the orientation  $\varphi_i$  of the particle 
is governed by the equations of motion:

\begin{equation}
 m_i\ddot{\vec{r}}_i  =\sum_{j}\vec F_{ji}-m_i g \hat y,  ~~~~~~
I_i\ddot{\varphi}_{i} =\sum_{j}{\tau_{ji}}. 
\label{eq:newton}
\end{equation}

\noindent 
Here $m_i$ and $I_i$ are the mass and moment of inertia of the particle. 
The sum is over all particles interacting with this particle; $g$ is the
gravity; and $\hat y$ is the unit vector along the vertical direction.

\subsection{Efficient calculation of forces} 
\label{efficient} 

The efficiency of the dynamics simulation is mainly determined  by the method of 
contact detection. In a system of $N$ particles, each one with $M$ edges, 
the number of operations required to update the positions of the particle 
in each time step is in the order of $O(N M)$, whereas the number of calculations for 
contact  detection is $O(N^2 M^2)$. Simulations therefore become very slow when 
either the number of particles or the number of vertices is large.

\begin{figure}[t]
  \begin{center}
    \epsfig{file= 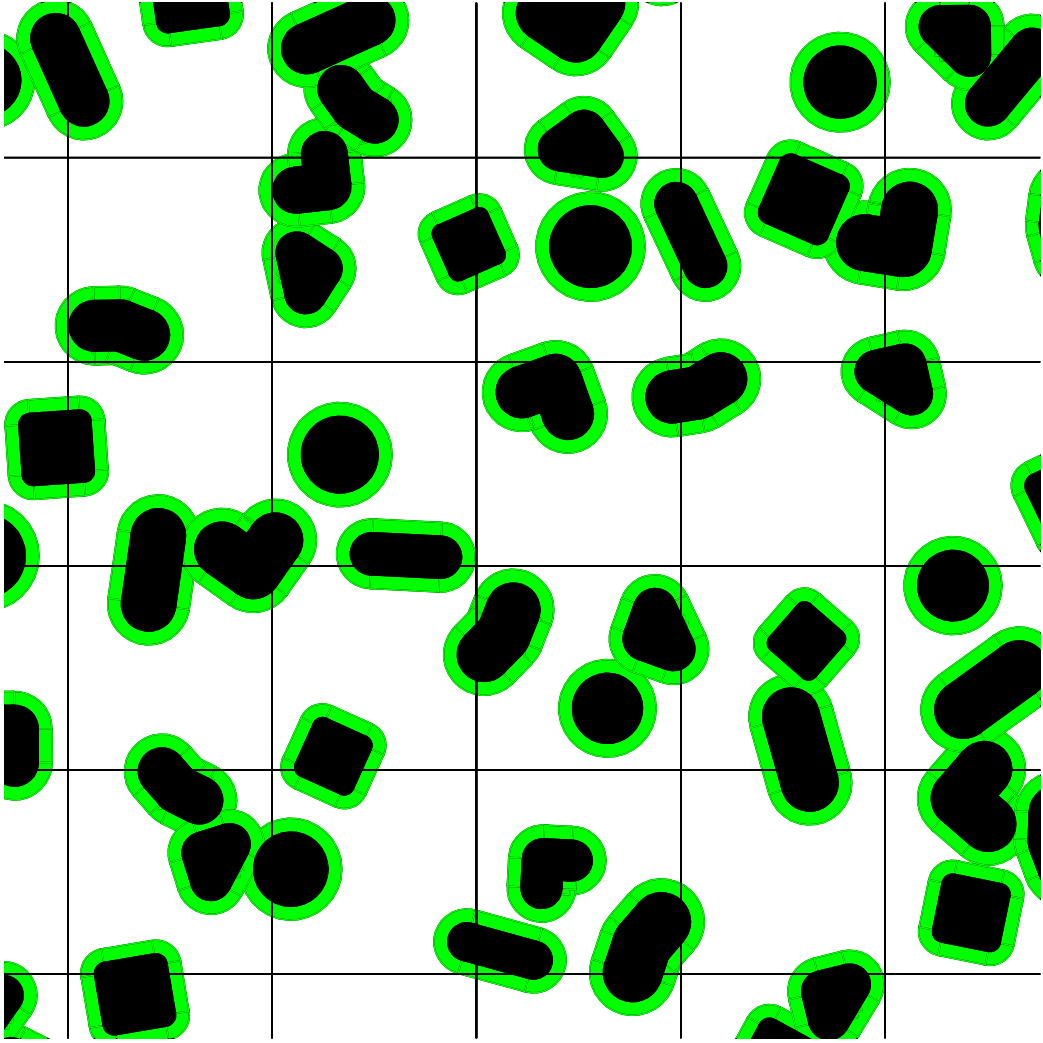,width=\linewidth}
   \caption{Method for identification of neighbor list: the space domain is divided by square cells.
Then the potential neighbor of the particles are those hosted in the same cells, or in the adjacent
cells.  Each particle has as {\it skin} of thickness $\delta$. If the skins of two potential neighbors 
overlap, they are included in the neighbor list.
}
   \label{fig:neighbor search}
  \end{center}
\end{figure}

The first step to speed up the simulations is to execute the force calculation only  over  
neighbor particles. With this aim we introduce the  {\it neighbor list}, which is the
collection of pair particles whose distance between them is less than  
$2\delta$. 
(The distance between two particles is defined as the minimum of all vertex-edge 
distances). The parameter $\delta$ is equivalent to the {\it Verlet distance}
used in simulations with spherical particles~\cite{poeschel04}. As shown in the 
Fig.~\ref{fig:neighbor search}, the Verlet method is equivalent to surround the
particles by a {\it skin}, so that the neighbors list consists of all particles
pairs whose skins overlap. 

A {\it link cell} algorithm \cite{poeschel04} is 
used to allow rapid calculation of this neighbor  list:   First, the space occupied by  
the particles is divided in cells  of side $D+\delta$, where $D$ is the maximal diameter of 
the particles. Then  the  link cell list is defined as the list of particles hosted in each cell. 
Finally, the candidates of neighbors for each particle are searched only in the  cell occupied by 
this particle, and its eight neighbor cells. 

The neighbor list is calculated at the beginning of the simulation, and it is 
updated when the following condition is satisfied:

\begin{equation}
 \max_{1 \le i \le N}\{\Delta x_i+R_i\Delta\theta_i\}>\delta.
\label{eq:neighbor update condition}
\end{equation}

\noindent
$\Delta x_i$ and $\Delta\theta_i$ are the maximal displacement and rotation
of the particle after the last neighbor list update. $R_i$ the maximal distance 
from the points on the particle to its center of mass.  After each update 
$\Delta x_i$ and $\Delta\theta_i$ are set to zero. The update condition is
checked in each time step. Increasing the value of  $\delta$ makes  updating 
of the list less frequent, but increases its size, and 
hence the memory  used in the simulation. Therefore, the parameter $\delta$ must 
be chosen by making a compromise between the storage (size of the neighbor list) and 
the computer time (frequency of list updates). 

Neighbor list reduces the amount of calculations to  $O(N M^2)$.  
Therefore the simulations are still very expensive when particles 
consist of a large number of vertices. Further reduction of the number of calculations 
between neighbor particles can be done by identifying which part of a particle 
is neighbor to the other. This idea is implemented as follows:  for each element
of the neighbor list, we create a {\it contact list}, which consists of those 
vertex-edge pairs whose distance between them is less than $r_i+r_j+2\delta$,
where $r_i$ and $r_j$ are the sphero-radii. 
In each time step, only these vertex-edge pairs are involved in the contact 
force calculations. Overall, neighborhood  identification requires a neighbor list 
with all pair of neighbor particles, and one contact list for each pair of neighbors. 
These lists require little memory storage, and they reduce the amount of calculations 
of contact forces to $O(N)$, which is of the same order as in simulations with 
spherical particles \cite{poeschel04}.

\subsection{Time Integration}
\label{time integration}
The equations of motion of the system are numerically solved using a four order 
predictor-corrector  algorithm \cite{poeschel04}. A pseudocode with the basic procedures 
in each time step is shown in Algorithm  \ref{alg: one step}.  The predicted method calculates the position 
(center of mass and orientation) of each particle and its derivatives using a Taylor expansion. 
Next the vertices of the all polygons are updated according to the predicted positions of the particles.
If the neighbor update condition of Eq.~\ref{eq:neighbor update condition} is satisfied, the link cell 
is calculated, and then it is used to update the neighbor list and the contact list of each one of its 
elements. Then the contact forces and torques are calculated. Finally, forces and torques are used to 
correct the position of the particles and their derivatives. The algorithm is basically the same as the 
used in polygons, except that the force is calculated using Eqs.~\ref{eq:contact force}.  
Note that the efficiency of the code is based in the simplicity of the contact force calculation, and 
in the fact that the Minkowski sum does not  need to be calculated during the time integration.

The parameters of the simulations are a constant stiffness $k=10^7 N /m $, 
gravity $g=10m/s^2$, density $\sigma=1kg/m^2$, time step $\Delta t = 10^{-5}s$ 
and Verlet distance $\delta = 1m$.

\begin{algorithm}[H]
\SetLine
\KwIn{state of the particles at time $t$}
\KwOut{state of the particles at time $t+\Delta t$ }

predict position of the particles and its derivatives\;
	
\If{neighbor update condition is satisfied}
    {  calculate link cell\;
       update neighbor list\;
       update contact lists\;
    }
update vertices of the particles\;
calculate contact forces between neighbor particles\;
apply contact forces to the particles\;
apply gravity forces to the particles\;
correct positions and its derivatives using forces and torques\;
\
\caption{One time step of the time integration scheme}
\label{alg: one step}
\end{algorithm}

\section{Non-dissipative granular dynamics simulations}
As a first step we present here simulations results of many body conservative systems.
We investigate the evolutions towards the statistical mechanical equilibrium of this
simple system. Generalization to dissipative system driven by external forces will
be presented in forthcoming papers.

\subsection{Energy balance}
The vertex-edge interaction between the particles is given by

\begin{equation}
\vec F(V,E) = k \delta(V,E) \vec N
\label{eq:elastic force}
\end{equation}

\noindent
where The material parameter $k$ is the stiffness constant,
$\delta(V,E)$ is given by Eq.~\ref{eq:overlap} and 
$\vec N$ is the unit normal vector:

\begin{equation}
\vec N = \frac{\vec Y-\vec X}{||\vec Y-\vec X||}
\label{eq:normal}
\end{equation}

Here  $\vec X$ is the position of the vertex $V$ and $\vec Y$ is its closest point 
on the edge $E$.

The question which now arises is whether the vertex-edge interaction in
Eq.~\ref{eq:elastic force}  leads to a  conservative system. Let us multiply the 
first equation in  ~(\ref{eq:newton}) by  $\dot{\vec{r}}$ and the second one by 
$\dot\varphi$. Next we sum both equations and then sum over all particles. 
After some algebra we get the energy conservation equation:

\begin{equation}
E_T=\sum_{ij}{E^e_{ij}}+\sum_i{(\frac{1}{2} m_i v^2_i + \frac{1}{2} I_i \omega^2_i)} = cte.
\label{eq:energy}
\end{equation}

\noindent
The first term of this equation corresponds to the potential elastic energy 
at the contacts:

\begin{equation}
E^e_{ij} =  \frac{1}{2} k (\sum_{V_i E_j}{\delta(V_i,E_j) } 
        +\sum_{V_j E_i}{ \delta(V_j,E_i)\ }).
\label{eq:potential}
\end{equation}

\noindent
The other terms of Eq.~(\ref{eq:energy}) are 
the linear and rotational kinetic energy of the particles. We emphasized 
that the elastic force in Eq.~(\ref{eq:contact force}) belongs to the potential 
energy  defined by Eq.~(\ref{eq:potential}), which proves that our model is 
conservative. The  simplicity of this force contrasts to the P\"oschel's model 
for interacting triangles \cite{poeschel04}, where the forces and torques associated 
to his potential energy lead to a much more expensive calculation. 

\begin{figure}
  \begin{center}
       (a)~~~~~~~~~~~~~~~~~~~~~~~~~~~~~~(b)\\
    \epsfig{file=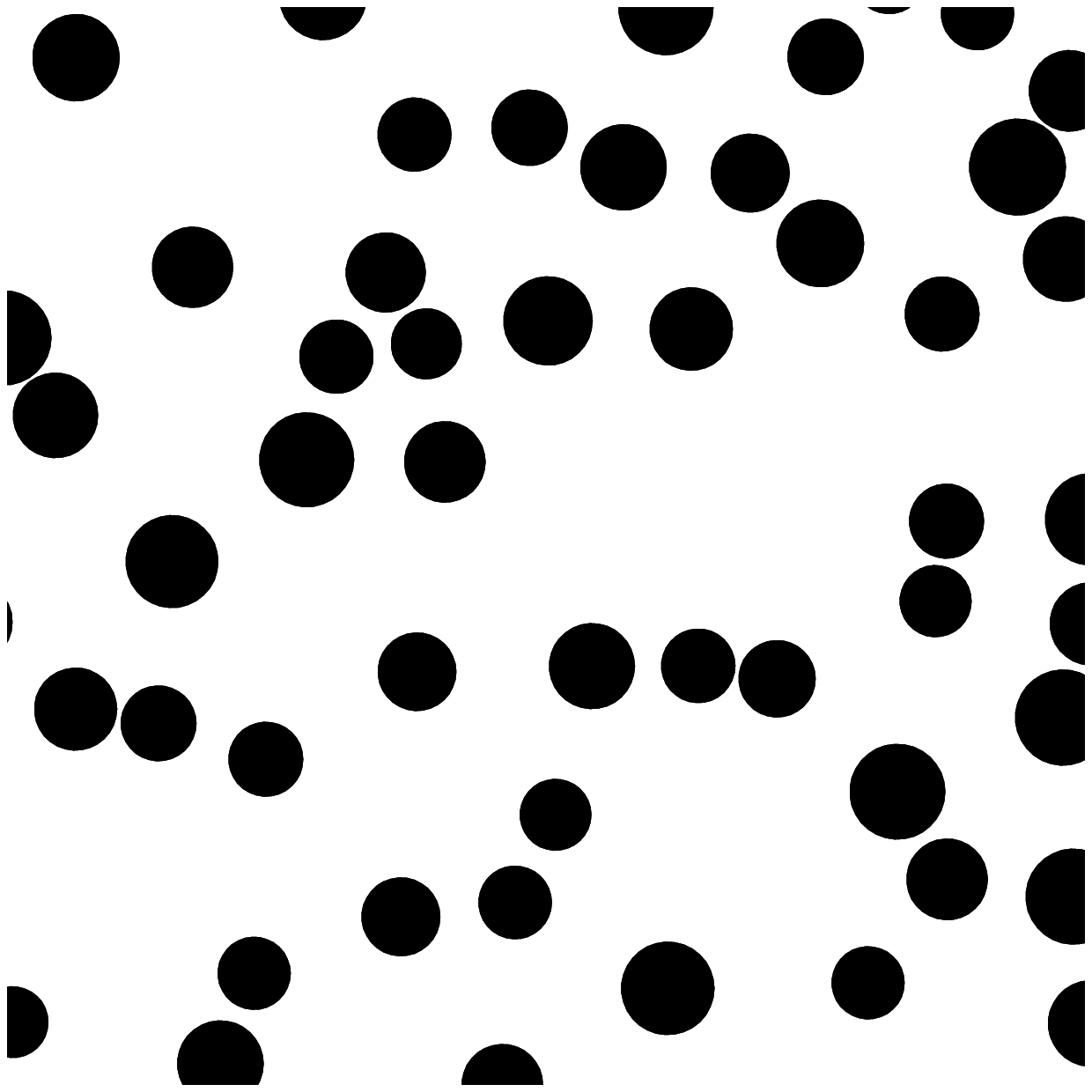,width=0.45\linewidth}~~~
    \epsfig{file=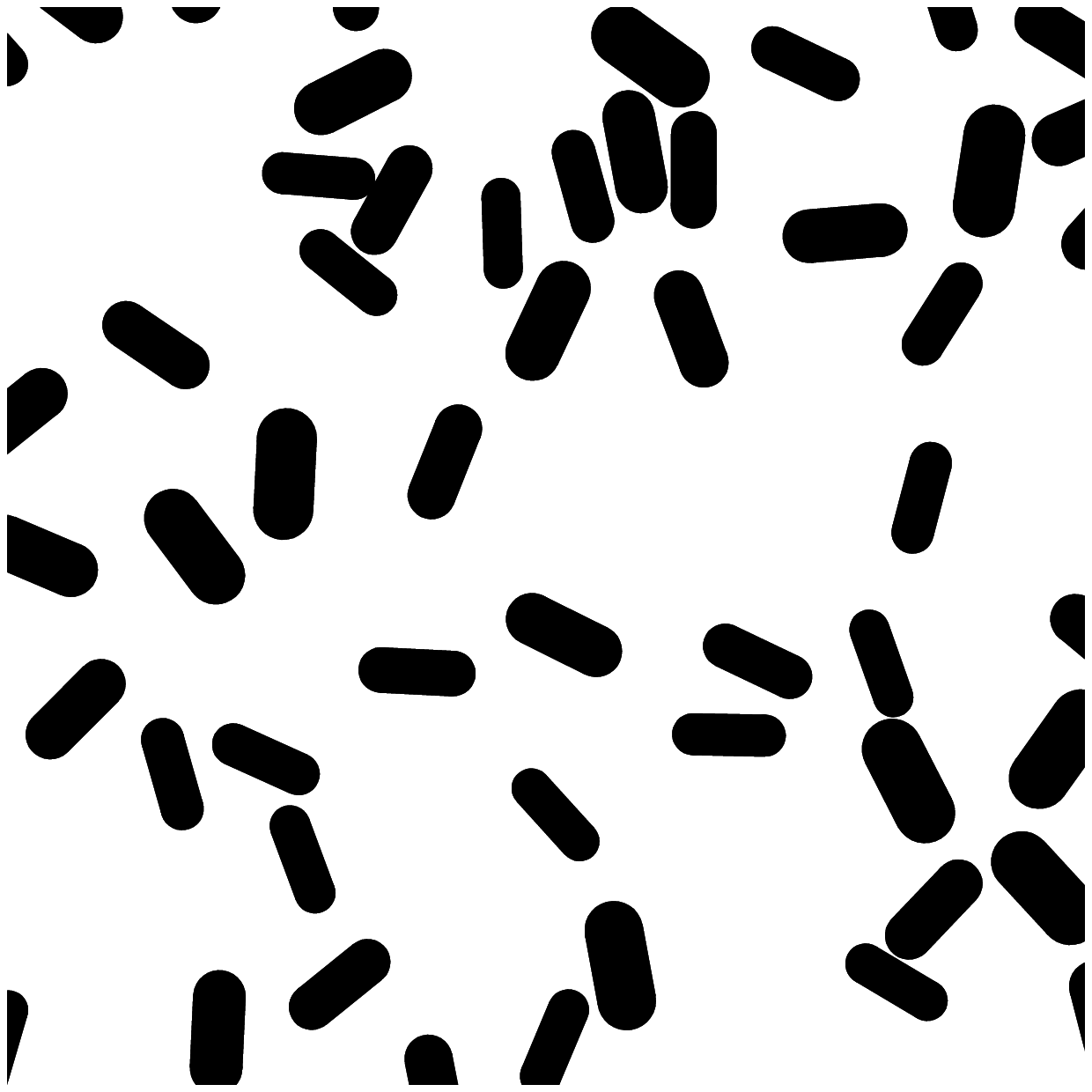, width=0.45\linewidth}
       (c)~~~~~~~~~~~~~~~~~~~~~~~~~~~~~~(d)\\
    \epsfig{file=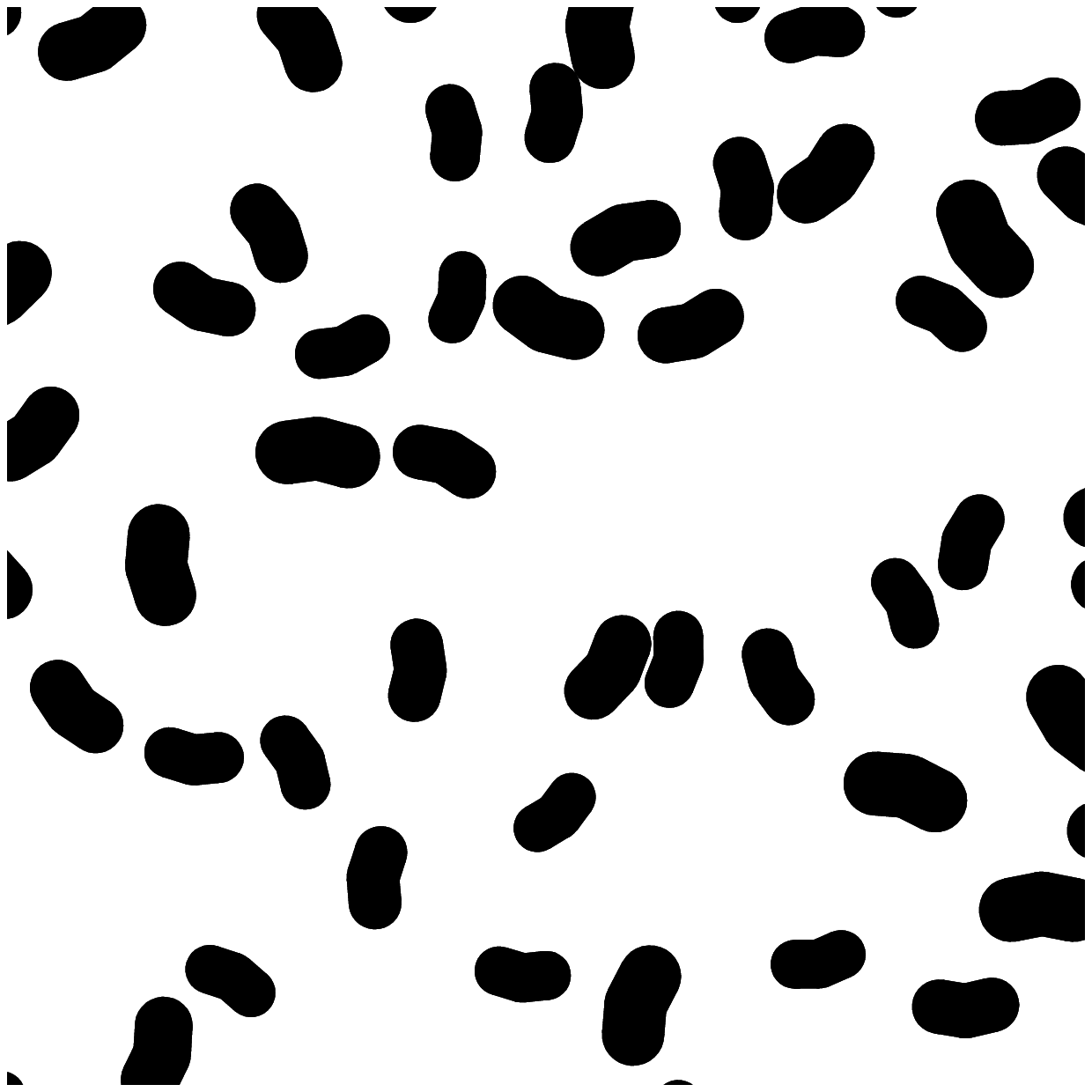, width=0.45\linewidth}
    \epsfig{file=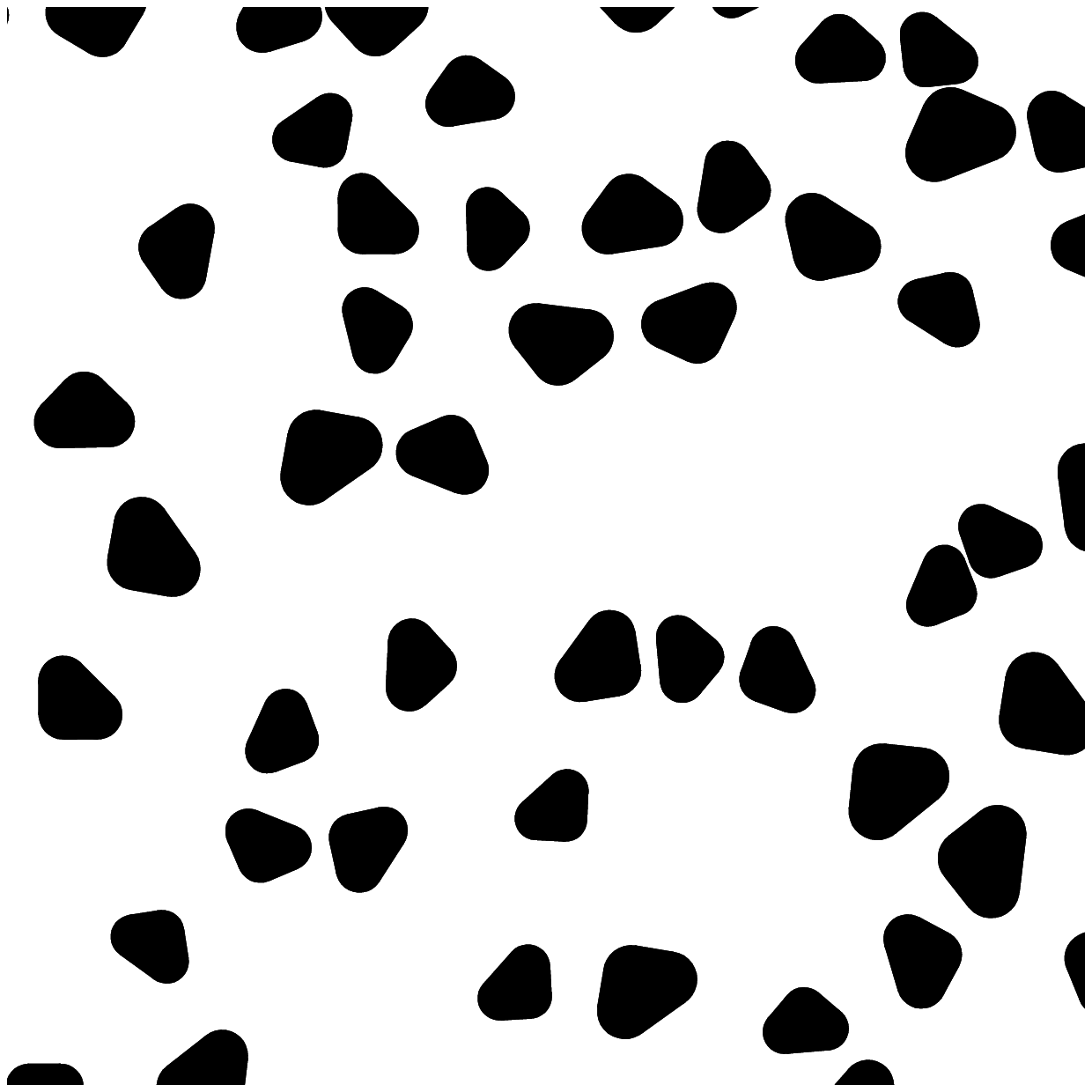, width=0.45\linewidth}
   \caption{
 Systems obtained from Minkowski sum approach:
(a) disks (Vertex + disk); (b) rice (segment + disk);
(c) peanuts (Polyline + disk) and (d) pebbles (triangle +disk)
}
   \label{fig:Minkowski sum}
  \end{center}
\end{figure}

Other important aspect of this dynamics simulation is the accuracy of the numerical 
solution.  The numerical error in the energy calculation is evaluated by performing 
a series of simulations with many  non-spherical particles interaction via the elastic
force given by Eq.~\ref{eq:elastic force}. Each test consists of $400$ particles 
confined by four fixed rectangular  walls. Each particle occupies an area of $1cm^2$ and 
the confining area is $46cm\times46.25cm$. These dimensions lead to a volume  fraction 
of $\Phi = 0.186$. Each sample  consists of identical particles  with a specific 
spheropolygonal shape  as shown   Fig.~\ref{fig:Minkowski sum}: disks (point+disks) 
rice (line+disk),  peanuts (polyline+disk), and pebbles (triangle+disk).

\begin{figure}[t]
  \begin{center}
    \epsfig{file=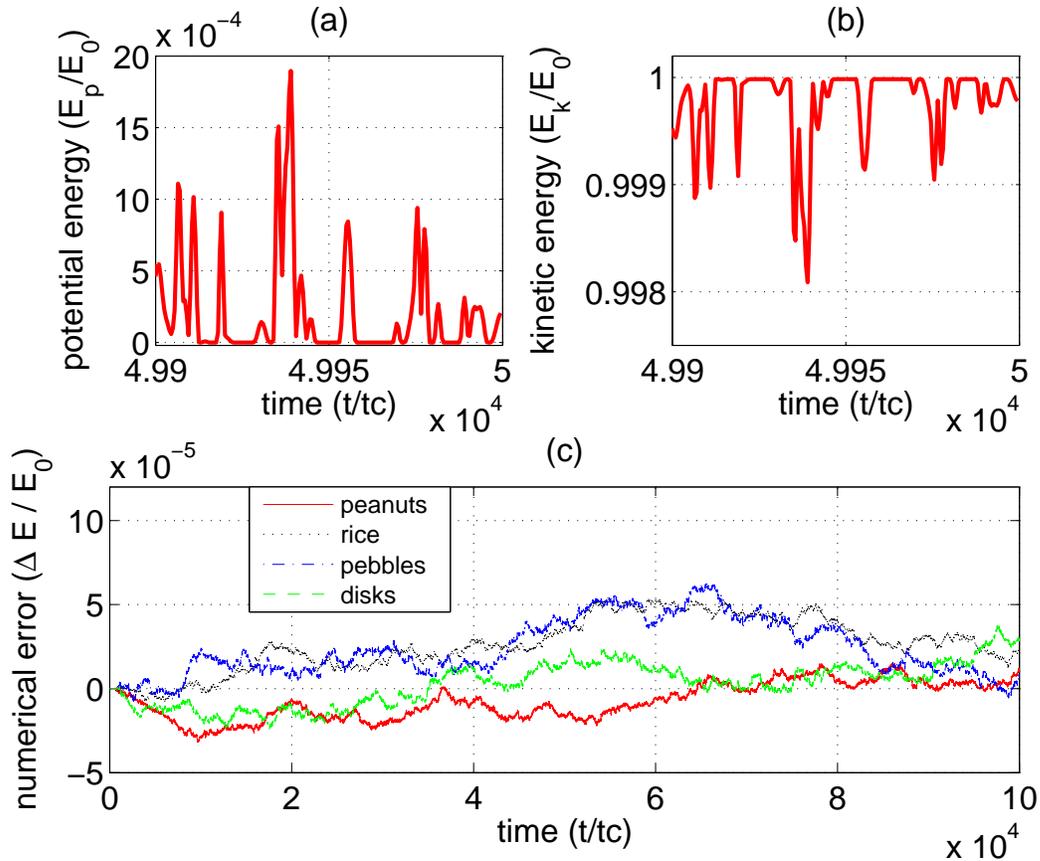,width=\linewidth}
   \caption{Time evolution of the total kinetic and potential energy in
the non-dissipative system.  As is expected from the energy conservation,
the total energy keeps constants, except numerical error wicho are lower than
$0.01\%$ thoughout all the simulations.
}
   \label{fig:energy}
  \end{center}
\end{figure}

Initially, each particle has zero angular velocity and a linear velocity of 
$1cm/s$ with  random orientation. Due to collisions,  the linear momentum of each 
particle changes and part of it is transferred to angular momentum. Fig.\ref{fig:energy}
show the potential (a) kinetic (b) and total (c) energy of the system. Elastic energy 
has a negligible contribution to the energy budget, as it differs from zero only for short 
times  during collisions. Energy conservation is numerically verified within a porcentual 
error of $0.01\%$. The energy fluctuations are produced by time discretization
We shall note alse that energy have a trend to grow slowly in all samples.

\subsection{Statistical equllibrium}

Simulations with a large number of particles show that elastic interactions
allow the particles to exchange energy and momentum, whereas its contribution to the 
energy budget is negligible. This property leads us to investigate the existence of 
a statistical equilibrium  in a gas of  non-spherical particles.  It is expected that 
the system will reach the statistical equilibrium, which is characterized by an
energy equipartition and a   Maxwell-Boltzmann statistics for energy  distribution 
\cite{tolman}:

The energy of the system consist of rotational and translational kinetic energy. At 
the beginning of the simulations the  rotational kinetic energy is zero, and it 
increases during the simulations due to collision, see left part of the  F
ig.~\ref{fig:statistical equilibrium}.   For all the samples, we observe the same 
stationary  regime, where  the average of   rotational kinetic energy reaches the 
limit of $1/2$  of the linear kinetic energy.   This is in agreement with the 
theorem of equipartition  of energy \cite{tolman},  which states that each 
quadratic term in the energy should  contribute the same  weight in the mean energy 
of the system.
.
It is expected that the system reach the statistical equilibrium, which is
characterized  by a  Maxwell-Boltzmann statistics for energy  distribution 
\cite{tolman}:

\begin{equation}
\rho(E_k)dE_k = 2\sqrt{E_k/\pi(kT)^3} \exp(-E_k/k_BT)dE_k,
\label{eq:Maxwell-Boltzmann}
\end{equation}

\noindent
here $E_k=\frac{1}{2}(mv^2_x+mv^2_y+I\omega^2)$ is the kinetic energy of the particle;
$T$ the temperature; and $k_B$ the Boltzmann constant. The mean energy of the 
particles leads to $\bar E_k=\int^\infty_0{\rho(E_k)E_kdE_k}=\frac{3}{2}k_B T$.

\begin{figure}[t]
  \begin{center}
    \epsfig{file=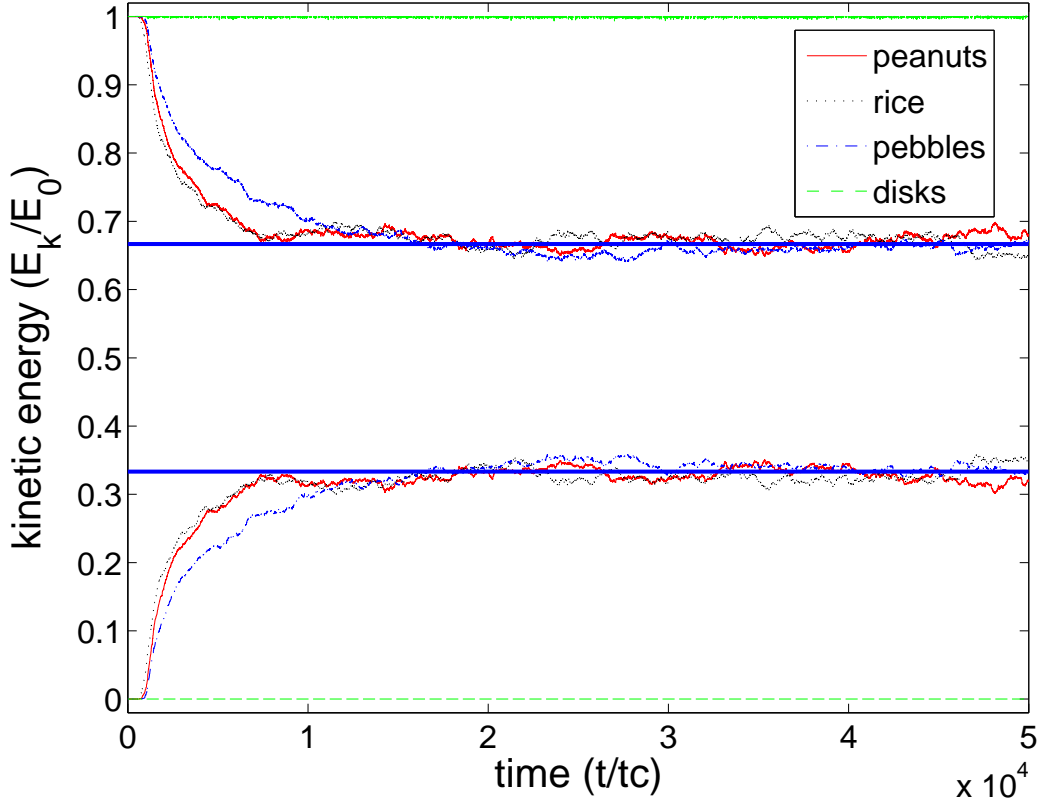,width=\linewidth}
   \caption{Time evolution of the total linear and rotational
kinetic energy of the particles. $E_0$ is the initial value of the total 
kinetic energy. The horizontal lines correspond to the expected value 
by the equilibrium statistical mechanics.}
   \label{fig:equipartition}
  \end{center}
\end{figure}

\begin{figure}
  \begin{center}
    \epsfig{file=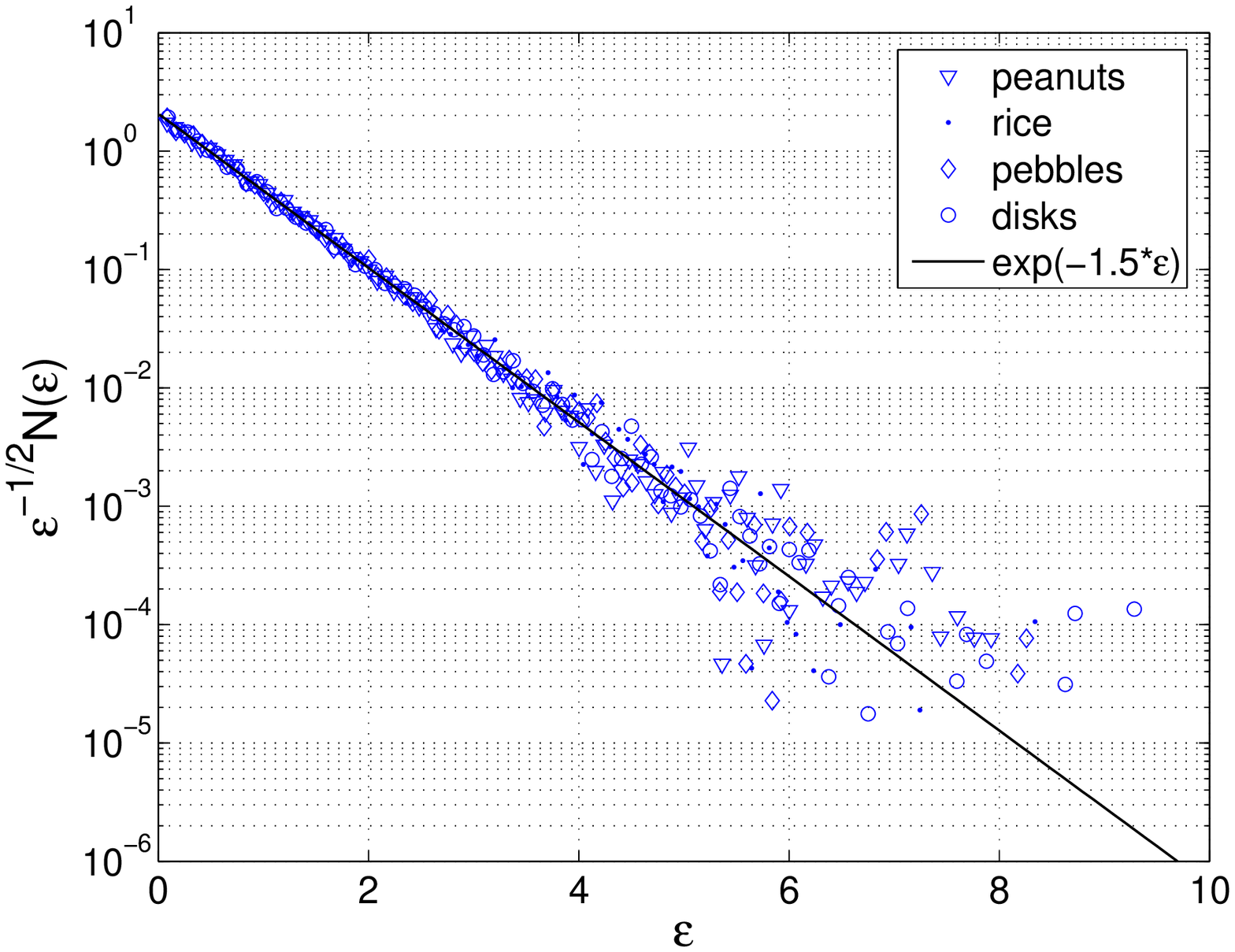,width=\linewidth}
   \caption{energy distribution $N(\epsilon)$ for the particles, where 
$\epsilon=E_k/\bar E_k$.The line corresponds to the best fit 
$n(\epsilon)= 2\beta \sqrt{\epsilon\beta/\pi} \exp(-\beta \epsilon)$,
with $\beta = 1.5$.
}
   \label{fig:energy distribution}
  \end{center}
\end{figure}

We also calculate the energy distribution of the particles in the
stationary regime. In order to calculate the energy distribution, we take 
snapshots of the simulations between $t=1s$ and $t=8s$ distanced by $0.01s$.  
In each snapshot the kinetic energy of the individual particles is measured. The 
histogram of the  variable $\epsilon = E_k/\bar E_k$ is constructed using $100$ identical bins between 
zero and the maximal value. According to  Eq.~(\ref{eq:Maxwell-Boltzmann}), the 
theoretical distribution of $\epsilon$ must satisfy
$\rho(\epsilon)d\epsilon = 2\sqrt{\epsilon \beta^3} \exp(-\beta \epsilon)d\epsilon$,
where $\beta = 3/2$. An excellent agreement between this theoretical distribution
and the simulations data is shown in Fig.~\ref{fig:energy distribution}. Simulations
with different non-spherical shapes, which will be presented elsewhere \cite{alonso08a}, 
show that energy distribution for all samples collapse onto the theoretical expected value. 
It is also shown that the relaxation time for the statistical equilibrium is very sensitive 
to the degree of non-sphericity of the particle.

\section{Dissipative granular dynamics simulation}

Here we present biaxial tests simultions with circular and non-circular particles.
Usually, the granular assemblies are compacted and loaded within a 
set of confining walls. These walls act as boundary conditions, 
and can be moved by specifying their velocity or the force applied 
on them. The response of the walls can be used to calculate the 
global stress and strain of the assembly.

The interaction of the spheropolygonspolygons with the walls is modeled here by 
using a simple visco-elastic force. First, we allow the polygons 
to penetrate the walls. Then, for each vertex of the polygon $\alpha$
inside of the walls we include a force

Confining walls can be used to generate samples with different void
ratios. Starting from a very loose packing, the sample is compacted by 
applying a centripetal gravitational field to the particles and on the 
walls, oriented to the center of mass of the assembly. Then the sample 
is subjected to an isotropic compression until the desired confining 
pressure is reached. In order to generate dense samples, the interparticle 
friction is set to zero during the construction. The loose samples are created 
taking damping coefficients 100 times greater than those used in the test 
stage. Samples with void ratio ranged from $0.128$ to $0.271$ 
can be achieved with this method \cite{pena07}.

We have investigated shear deformation of granular samples with 
different initial void ratios~\cite{pena07}. Shear bands are observed in 
dense samples, whereas they seems to be absent in loose ones. they share 
some common properties of the shear bands in real granular materials,
such as their characteristic reflection when they reach the boundary wall.  
Shear band orientation lies between the Roscoe angle and Mohr-Coulomb 
solution, as in most of the experimental data.

For large shear deformations all samples reaches the critical state, which
is independent on their initial density. Once the samples reaches this 
state, they deform  at constant void ratio and coordination number~\cite{pena07}. The 
evolution of  the deviatoric stress exhibits fluctuations around the residual 
strength. Abrupt reduction of the stress results from the collapse of 
force chains, as shown the Fig.\ref{Fig:Collapse_fch}. collapse of 
force chains makes the sample to approach and retreats unstable
stages. A similar  behavior is observed  in glass bead samples 
\cite{nasuno97} and packings of 
glass spheres \cite{adjemian05}. Experimental biaxial tests show evidence of 
\textsl{dynamic instabilities} at the critical state \cite{vardoulakis05}. 
Erratic slip-stick motion at the critical state is interesting, owing to  
its potential analogy with earthquake dynamics \cite{alonso06}.

\begin{figure}
  \begin{center}
       (a)~~~~~~~~~~~~~~~~~~~~~~~~~~~~~~(b)\\
    \epsfig{file=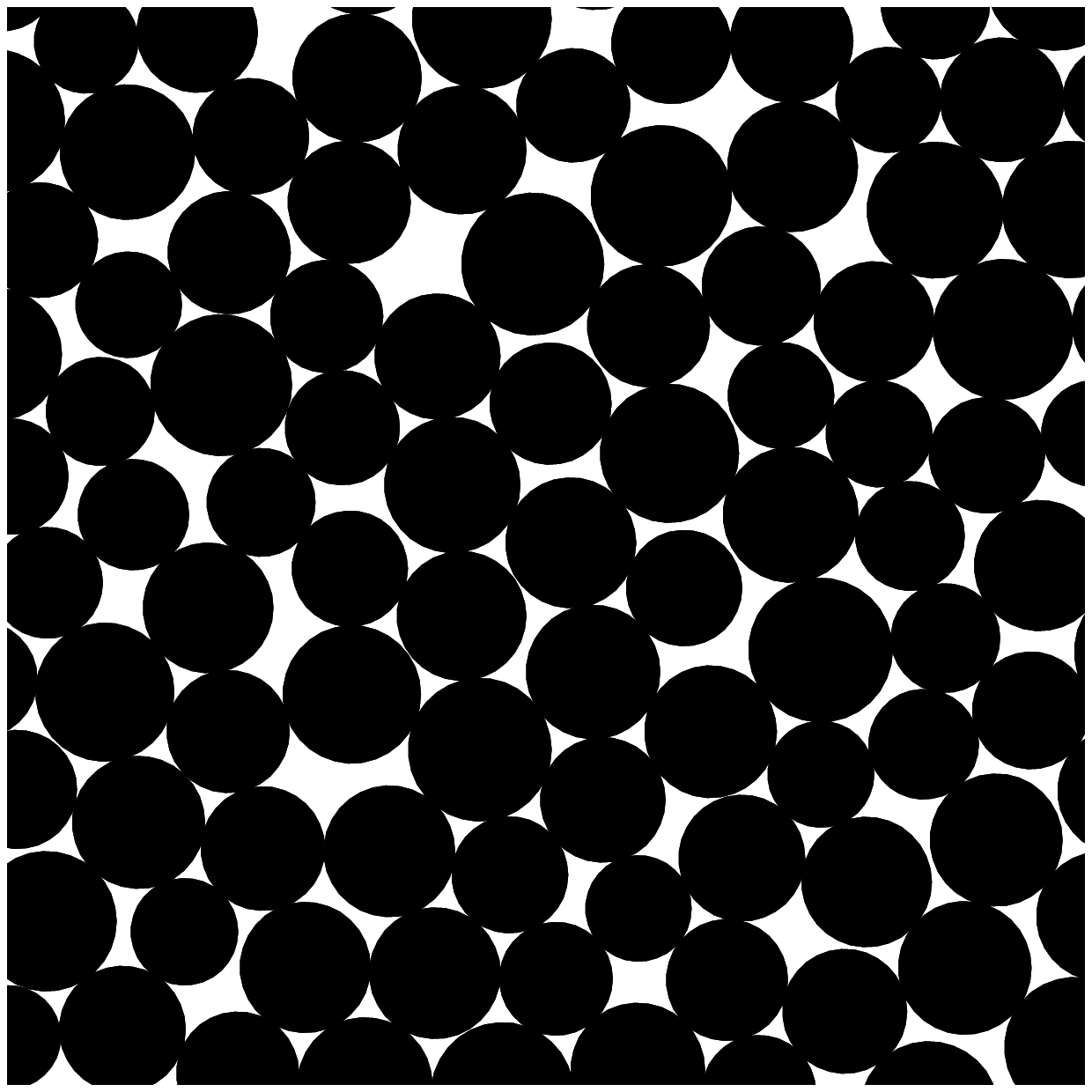,width=0.45\linewidth}~~~
    \epsfig{file=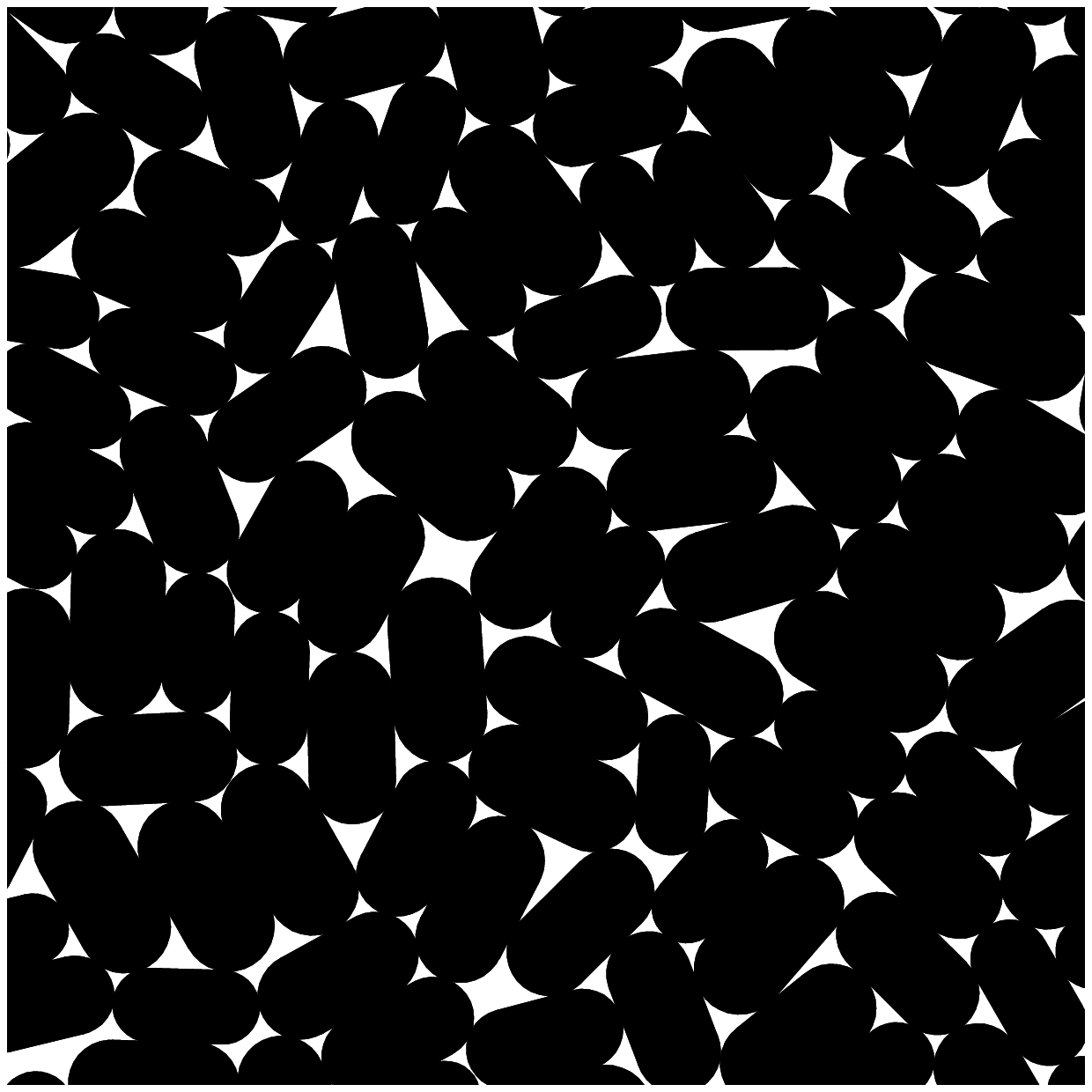, width=0.45\linewidth}
       (c)~~~~~~~~~~~~~~~~~~~~~~~~~~~~~~(d)\\
    \epsfig{file=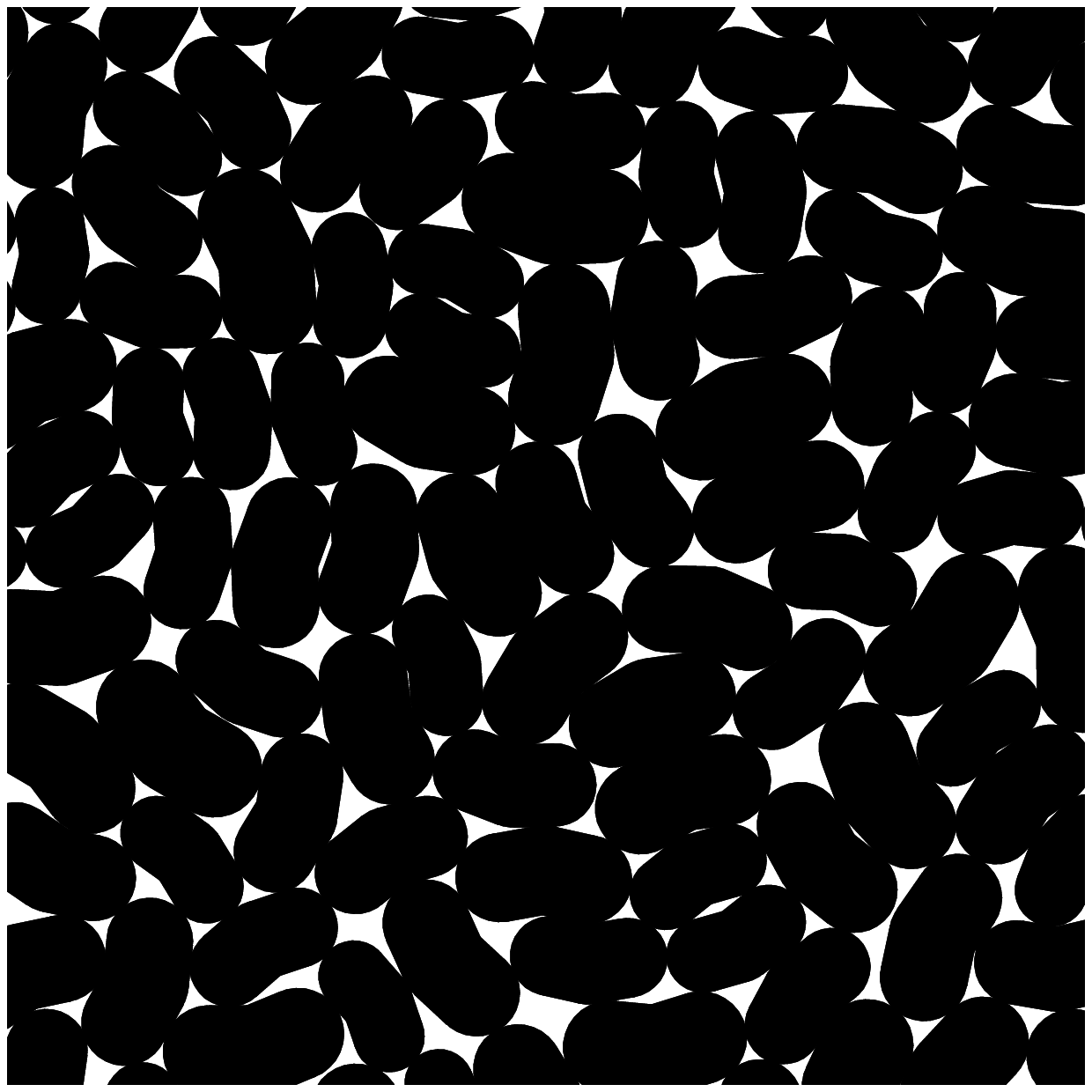, width=0.45\linewidth}
    \epsfig{file=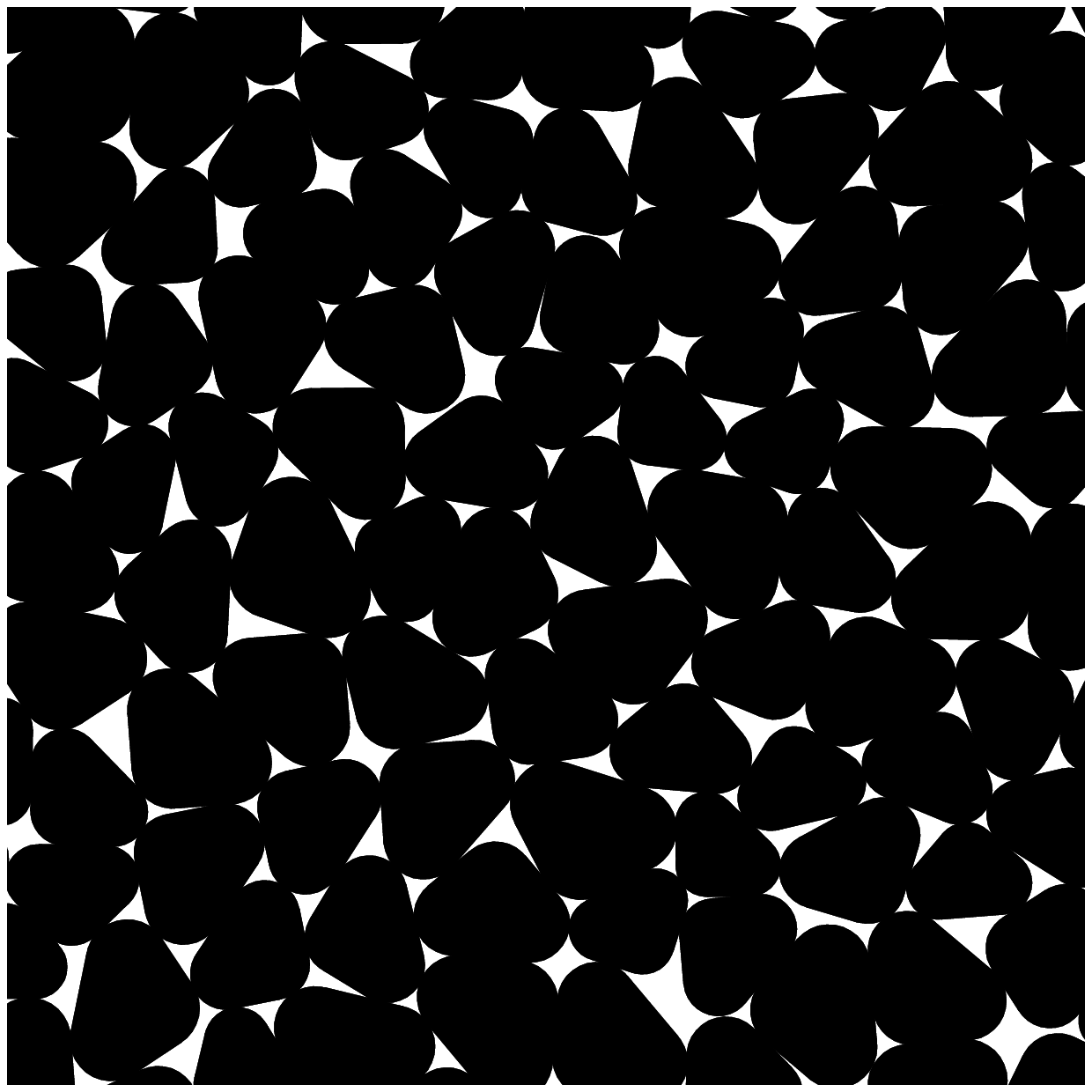, width=0.45\linewidth}
   \caption{
Granular packing obtained with disks, rice, peanuts and rice
}
   \label{fig:Minkowski cows}
  \end{center}
\end{figure}

\section{Performance}
Lastly, We compare the efficiency of the many-body simulations of systems consisting 
on disks, spheropolygons  and clums of spheres. Each spheropolygons correspond to a 
particle with complex shape, and it consist on $62$ vertices.The clump of spheres represents 
the same complex shaped particles, and it need $726$ particles. The macroparticles 
they are simulating by summing up the contact forces between the constituting disks, and  updating all the disks of 
each particle according with its current position.
The performance of the simulations is estimated by  running different processes in a 
Pentium 4, 3.0GHz, and calculating  the  {\it Cundall number} is each one of them. 
This number is the amount of particle time  steps executed by the processor in one second, 
which is calculated as  $c = N_T N /T_{CPU}$, where $N_T$ is the number of time steps, $N$ is the 
number of particles and $T_{CPU}$ is the CPU time of the simulation.  
Fig.~\ref{fig:performance}~shows the Cundall number versus the
number of particles for the three  cases. 
When the number of particles is between $N = 100$ and $N=1000$ the Cundall number is 
approximately constant. This constant  is around  $100,000$ for disks, 
$2,000$ for spheropolygons, and $50$ for clumpy particles. Therefore the simulations
with spheropolygons, although slower than simulations with disks, are much
faster than simulations with clumpy particles.  This is because each time step needs to 
update the position of  $62$ vertices  of the shperopolygons  whereas it needs to 
update the position of the $726$ disks in the  case of the clumpy particles.  Therefore 
simulations with spheropolygons are more efficient than those ones with clumps of 
disks, because the former ones require less elements to represent the particle 
shape.

\begin{figure}
  \begin{center}
    \epsfig{file=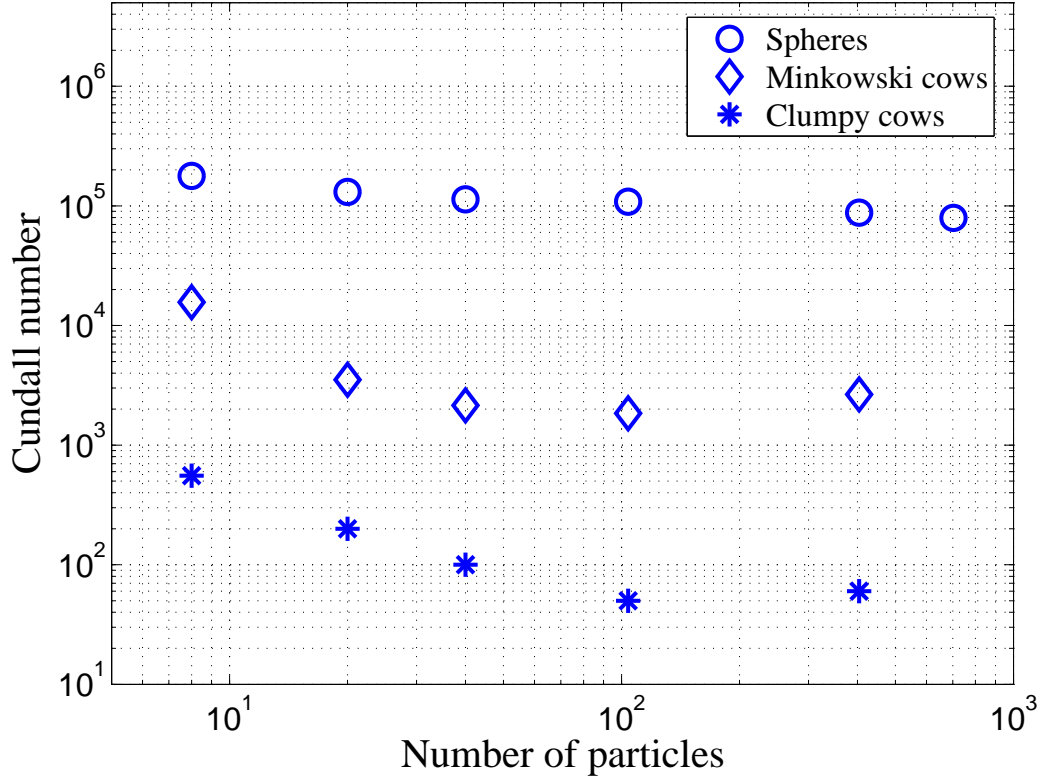,width=\linewidth}
   \caption{Cundall number versus the number of particles, in simulations with
disks, spheropolygons and clumpy particles.
}
   \label{fig:performance}
  \end{center}
\end{figure}

\begin{figure}
  \begin{center}
      (a)\\
    \epsfig{file=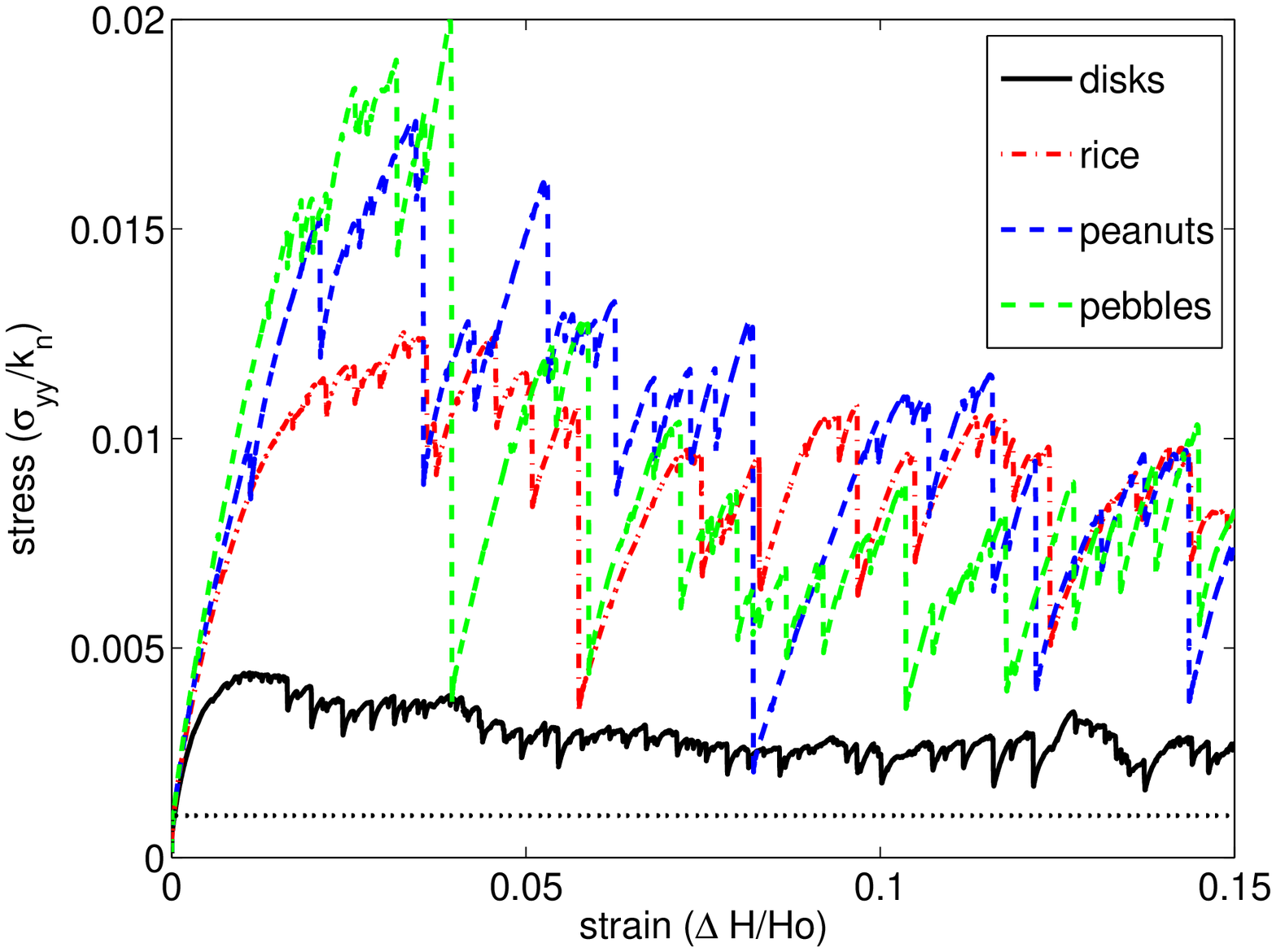,width=\linewidth}\\
    (b)\\
    \epsfig{file=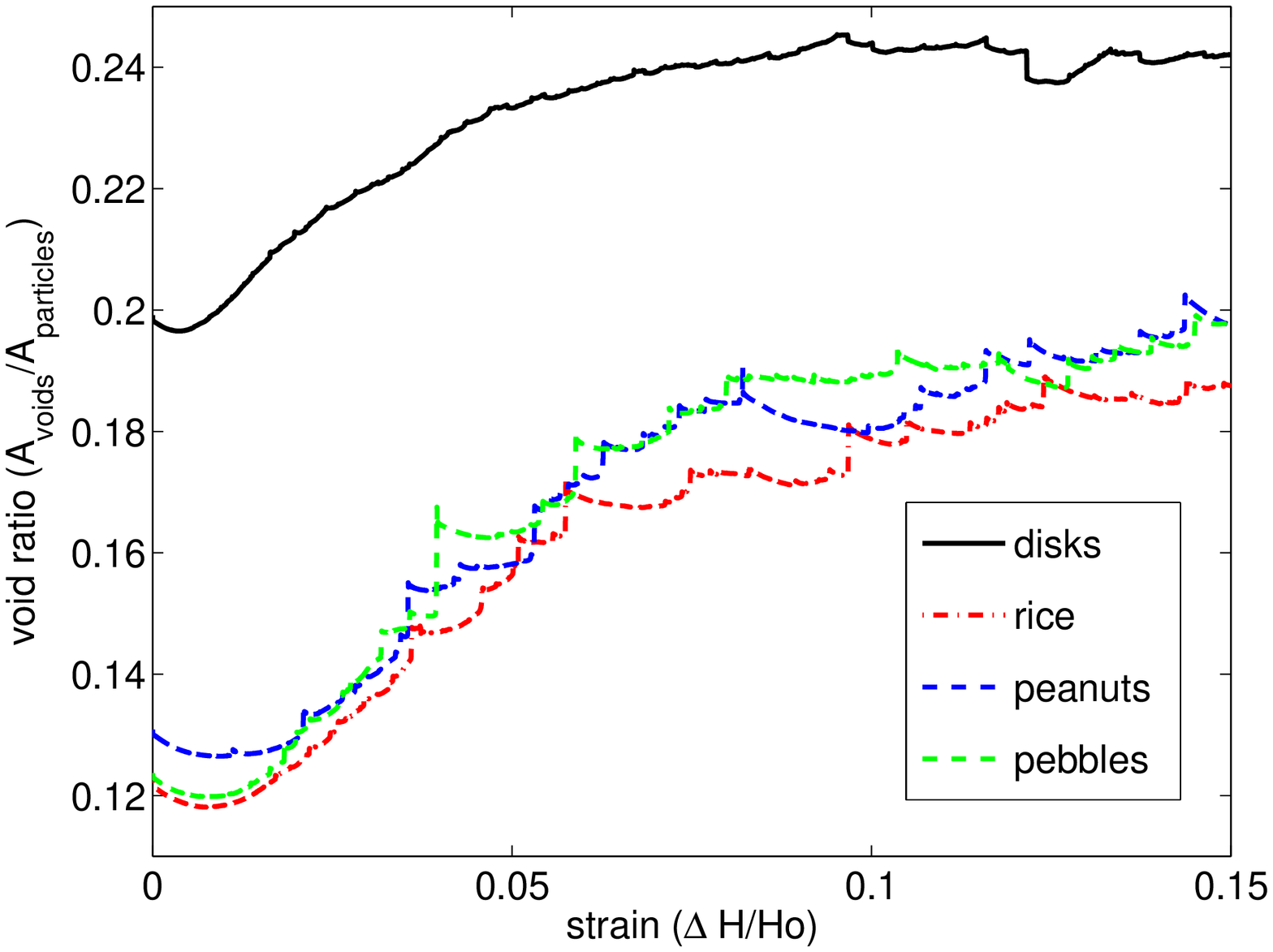,width=\linewidth}
   \caption{Stress versus strain (a) and void ration versus strain (b) for different 
            particles shape, in biaxial test simulations.The dotted line in (a) represents
            $p/k_n$, where p is the applied pressure on the latteral walls.}
   \label{fig:performance}
  \end{center}
\end{figure}

\section{Concluding remarks}

The  method presented here provide an energy balance equations and a wide 
range of particle shape representations, 
including non-convex particles and tunable grain roundness.
This paper shows that modeling interacting particles using spheropolygons has 
several advantages  with respect to other existing particle-based models: 
i) The possibility to model non-convex particles; ii) a realistic representation of the surface 
curvature of particles; iii) guaranteed compliance with physical and 
statistical mechanical laws; iv) balance between accuracy and efficiency. 
Benchmark tests prove energy conservation with an error below $0.0001\%$. 
Simulations with many particles verify the Maxwell-Boltzmann distribution and the principle
of energy equipartition. The computational efficiency is compared to simulations with disks
and clumps of disks. Simulations with disks are around $50$ times faster than simulation 
with spheropolygons.  However, the speed of the simulation is $40$ times faster that 
simulations with clumpy particles.

THe method overcome also two main difficulties existing in previous DEM developments.

(1) The interaction between polygons using the overlapping area is 
difficult to generalize in 3D, because  the overlap between polyhedrons 
is much more difficult to evaluate.

(2) The elastic force used in this work does not belong from a 
potential, so that this model does not provide an equation for 
energy balance. In the investigation of fault zones, the energy 
balance is required  to determine the energy  budget in earthquakes.

The model is still very simple, but extensions to more complex interactions
and 3D simulations are achievable in the near future. Cohesive and frictional forces 
can be incorporated by introducing internal variables in each vertex-edge contact.
These variables account for elastic deformation at the contacts and they must be
updated in each time step according to the sliding conditions or breaking criterion. 
3D modeling using spheropolyhedra requires elastic forces similar to 
Eq.~(\ref{eq:contact force}), where the sum is over all vertex-face  and edge-edge 
interactions. Special attention is required  for the case of two parallel edges in contact. 
This case lead to a non-uniqueness in the  selection of contact points. This need to be 
resolved to get a physical correctness in the torque calculation.

I thank Syed Imran for technical support;
R. Cruz-Hidalgo and K. Steube for review of an early version of 
the manuscript; A.J. Hale and M. L. Kettle for writing corrections;
and S. Latham, Weatherley, E. Heesen,  H. Muhlhaus,  P. Mora, W. Hancock, 
P. Clearly, S. Luding, and S. McNamara for  discussions. This work is supported by
the Australian Research Council (project number DP0772409 ) and the AuScope project.

\bibliographystyle{elsart-num}

\bibliography{/home/fernando/projects/references2}

\end{document}